\documentclass[]{aa}  
\input epsf.tex
\title{ISO far-infrared observations of rich galaxy clusters\thanks{
Based on observations with ISO, an ESA project with instruments founded by
ESA member states (especially the PI countries: France, Germany, the
Netherlands, and the United Kingdom) and with the participation of ISAS
and NASA}
}
\subtitle{II. \object{S\'{e}rsic\,159-03}}
\author{L. Hansen\inst{1} 
\and H.E. J{\o}rgensen\inst{1}
\and H.U. N{\o}rgaard-Nielsen\inst{2}
\and K. Pedersen\inst{2}
\and P. Goudfrooij\inst{3,4} 
\and M.J.D. Linden-V{\o}rnle\inst{1,2}
}
\institute{
Copenhagen University Observatory,
Juliane Maries Vej 30,
DK-2100 Copenhagen, Denmark
\and 
Danish Space Research Institute,
Juliane Maries Vej 30,
DK-2100 Copenhagen, Denmark
\and 
Space Telescope Science Institute,
3700 San Martin Drive, Baltimore, 
MD 21218, USA 
\and
Affiliated to the Astrophysics Division, Space Science Department,
European Space Agency}
\offprints{L. Hansen}
\mail{leif@astro.ku.dk}
\authorrunning{Hansen et al.}
\titlerunning{S\'{e}rsic\,159-03}
\date{Received date; accepted date}

\begin{document}
\thesaurus{03                         
(11.03.4 S\'{e}rsic\,159-03 
           13.09.1)                   
          } 
\maketitle
\begin{abstract}
In a series of papers we investigate far-infrared emission from rich galaxy 
clusters. Maps have been obtained by ISO at $60\mu {\rm m}$, $100\mu {\rm m}$, 
$135\mu {\rm m}$, and $200\mu {\rm m} $ using the PHT-C camera. Ground based
imaging and spectroscopy were also acquired. Here we present the results
for the cooling flow cluster S\'{e}rsic\,159-03. An infrared source coincident
with the dominant cD galaxy is found. Some off-center sources are also
present, but without any obvious counterparts.
\end{abstract}
\section{Introduction}
\label{introduction}
The first paper in this series (Hansen et al. \cite{han99}, paper{\sc \,i})
presented infrared data for the Abell\,2670 cluster. We identified 3 
far-infrared sources apparently related to star forming galaxies in the
cluster. The present paper concerns the rich cluster S\'{e}rsic\,159-03. 
The central part of the S\'{e}rsic\,159-03 cluster was mapped by the
Infrared Space Observatory (ISO) satellite, using the PHT-C camera 
(Lemke et al. \cite{lem96}) at $60\mu {\rm m}$, $100\mu {\rm m}$,
$135\mu {\rm m}$, and $200\mu {\rm m} $. The observations were performed
twice with slightly different position angles which gives an opportunity 
to do independent detections and to study possible instrumental effects.

The S\'{e}rsic\,159-03 cluster (Abell S1101, z=0.0564) is of richness class 0, 
Bautz-Morgan type{\sc \,iii} with a central dominant cD galaxy 
(Abell et al. \cite{abe89}). A cooling flow is 
present, and Allen and Fabian (\cite{all97}) found a mass deposition rate of
${\rm \dot{M} = 231^{+11}_{-10} ~M_{\sun} ~yr^{-1}}$ from ROSAT PSPC data. 
The cooling flow is centered on the cD~galaxy which exhibits nebular
line emission. Crawford and Fabian (\cite{cra92}) obtained optical spectra
and found from line-ratio diagrams that the ratios obtained along the slit
bridged the gap between class{\sc \,i} and class{\sc \,ii} in the scheme of 
Heckman et al.  (\cite{hec89}). Their spectra had position angle $90\degr$. 
West of the center they discovered a detached
filament of emission having extreme class{\sc \,ii} characteristics. They
argued that the different line ratios are due to changes in ionization
properties. Below in Fig.~\ref{emission} we show the  extent of the 
nebular emission.
In a subsequent paper  Crawford and Fabian (\cite{cra93}) included IUE data
to obtain the optical-ultraviolet continuum. They announced that a strong
Ly$\alpha$ line is present in the IUE spectrum.

\section{Observations}
\label{observations}
\subsection{The ISO data}
\label{ISOobs}
A rectangular area centered on the cD galaxy of S\'{e}rsic\,159-03
was mapped by ISO May 7, 1996 on revolution 173. The projected Z-axis of the
spacecraft had a position angle of $54 \fdg 4$ on the sky (measured from north 
through east). The observation
was repeated June 4, 1996, during revolution 200, but this time with position 
angle $69 \fdg 5$. The observing mode was PHT\,32 as for Abell\,2670
(paper{\sc \,i}). The 9 pixel C100 detector was used for 
$60\mu {\rm m} $ and $100\mu {\rm m} $ to map an area of
$10 \farcm 0 \times 3 \farcm 8$. For $135\mu {\rm m} $ and 
$200\mu {\rm m} $ the 4 pixel C200 detector was applied to cover a mapped 
area of $11 \farcm 0 \times 4 \farcm 6$. The target dedicated times were
1467 seconds for C100 and 1852 seconds for C200. 

\begin{figure}
  \epsfxsize=8.7cm
  \epsfbox{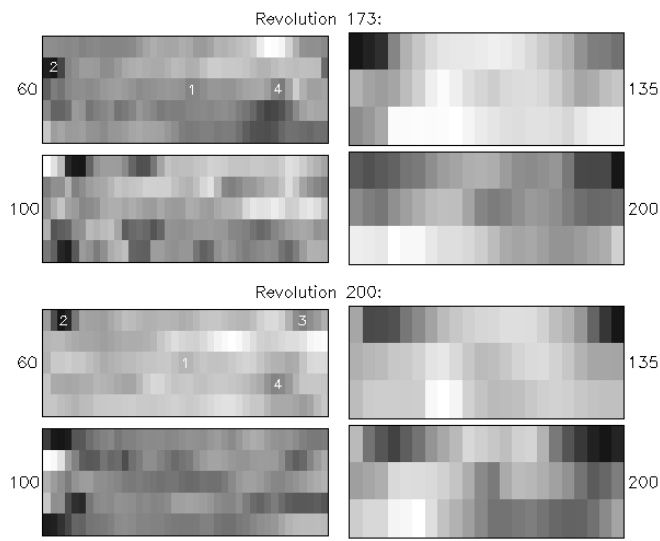}
  \caption[]{
The brightness maps for the four pass-bands are shown for revolutions 173 
and 200. Maximum brightness is dark. The maps are all centered on 
the dominant cD galaxy, but the revolution 200 maps are rotated $15 \degr$
counter-clockwise on the sky with respect to the revolution 173 maps. 
The C100 maps (left) cover $10 \farcm 0 \times 3 \farcm 8$ while the C200
maps (right) cover $11 \farcm 0 \times 4 \farcm 6$.  The features
marked with numbers in the $60\mu {\rm m} $ maps are regarded as real
sources. An optical image of the field is shown in Fig.~\ref{field}
             }
  \label{PIAim}
\end{figure}
\begin{table*}
\caption[]{
Parameters describing the spectroscopy
}
\begin{flushleft}
\begin{tabular}{l l l l l l }
\hline
Grism & P.A. & exp. & slit & $\Delta \lambda$ & range \\
     &       & min. & arcsec & \AA & \AA \\
\hline
\#8  & $21\degr$   & 40 & 1.5 &  4 & 5900-8300 \\
\#8  & $270\degr$  & 40 & 1.5 &  4 & 5900-8300 \\
\#10 & $21\degr$   & 45 & 1.5 & 24 & 3500-8800 \\
\#10 & $270\degr$  & 30 & 1.5 & 24 & 3500-8800 \\
\hline
\end{tabular}
\end{flushleft}
\label{spectro}
\end{table*}
As described in paper{\sc \,i} we apply the ISOPHOT Interactive Analysis 
software\footnote {The ISOPHOT data presented in this paper was reduced using 
PIA, which is a joint development by the ESA Astrophysics Division and the 
ISOPHOT consortium.} (PIA) for the reduction work. We also perform parallel
reductions using our own least squares reduction procedure 
(LSQ, cf. paper{\sc \,i}).
Although LSQ does not use sophisticated methods to correct for various 
effects -- e.g. glitches from cosmic rays are simply discarded -- we find 
it valuable for comparisons with the PIA reductions when evaluating the 
reality of features visible in the frames. The conclusion is that the PIA 
reduced images presented here (Fig.~\ref{PIAim}) do not contain noticeable 
artifacts from glitches. As in paper{\sc \,i} we present the data maps 
with pixel sizes $15 \arcsec \times 46 \arcsec$ for C100 and 
$30 \arcsec \times 92 \arcsec$ for C200, but the instrumental resolution
is only about $50 \arcsec$ for C100 and $95 \arcsec$ for C200 (paper{\sc \,i}).
The uncertainty of the maps increases towards the left and right borders due 
to the way the mapping was performed.

\subsection{Optical data}
\label{opt_obs}
\begin{figure}
  \epsfxsize=8.7cm
  \epsfbox{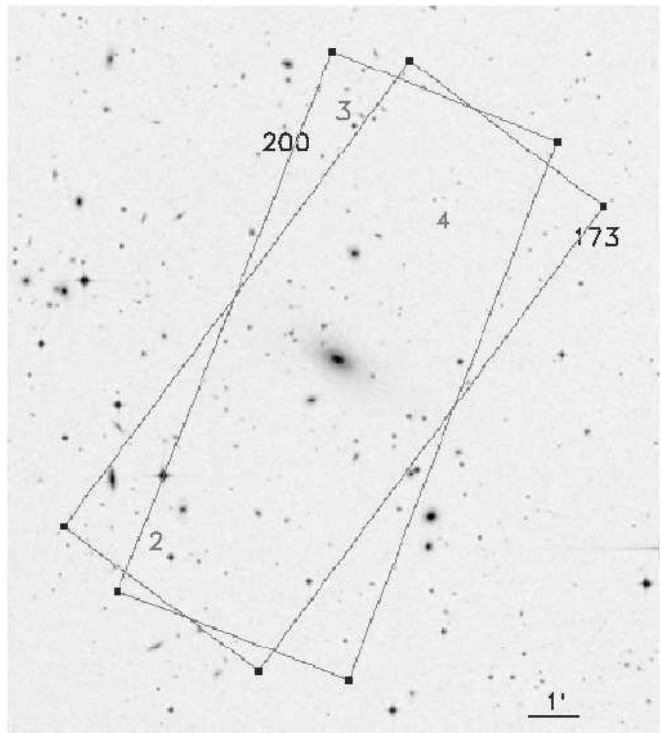}
  \caption[]{
%
%
An optical image of the central part of S\'{e}rsic\,159-03. North is up and 
east to the left. The areas covered by the two C200 mappings  are shown. 
For comparisons the map shown by Fig.~\ref{PIAim} should be rotated 
counter-clockwise by $54 \fdg 4$ for revolution 173 and $69 \fdg 5$ for
revolution 200. The numbers mark the approximate positions of off-center
sources
             }
  \label{field}
\end{figure}
\begin{figure}
  \epsfxsize=8.7cm
  \epsfbox{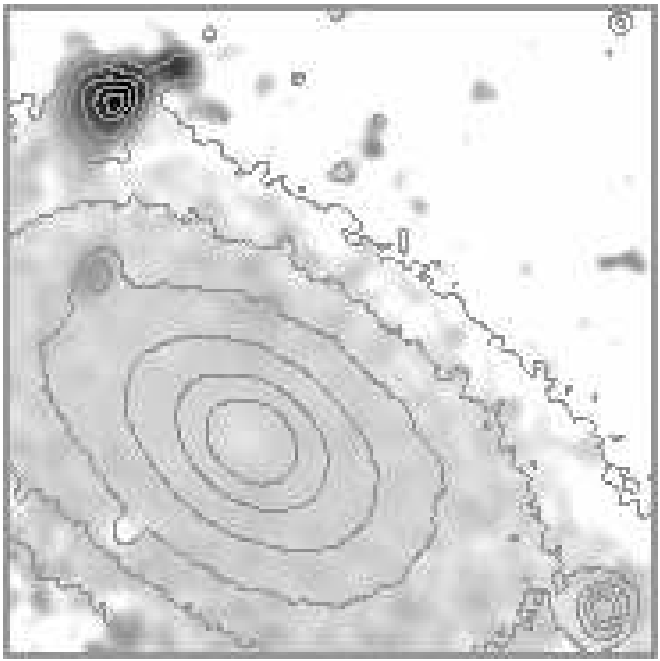}
  \caption[]{
The B-I colour distribution of the central part of the cD galaxy with
contours of the optical surface brightness overlayed (logarithmic scale). 
Dark depicts the
bluest colour. The upper left object is bluer than the cD galaxy by
$\Delta $(B-V)=-0.61 or $\Delta $(B-I)=-1.17. The field is 
$70\arcsec \times 70\arcsec$. North is up and east to the left
             }
  \label{B_I}
\end{figure}
\begin{figure}
  \epsfxsize=8.7cm
  \epsfbox{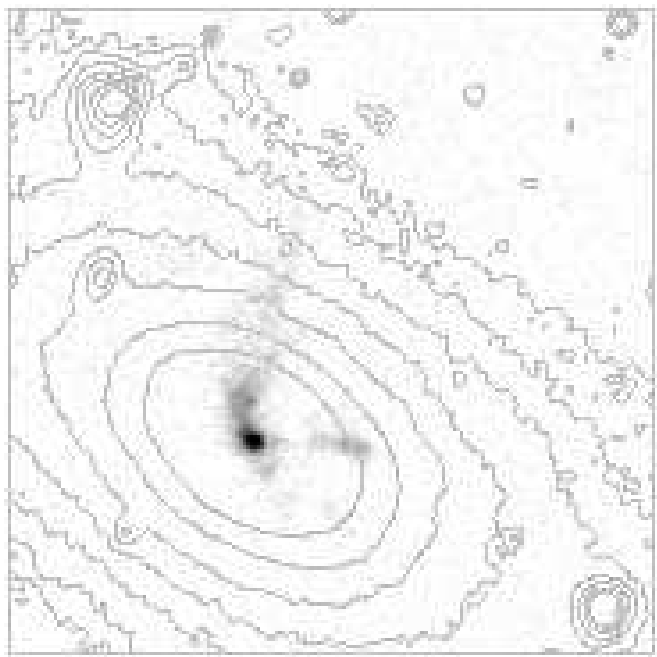}
  \caption[]{
The image of H$\alpha$+[N\,{\sc ii}] nebular emission overlayed with
contours of the optical image. Same field as Fig.~\ref{B_I}. 
A filament of emission points from the nucleus along
position angle $\approx 20\degr$ flaring towards north some $5\arcsec$
from the center. Emission towards the southwest is also seen. The filament
discovered by Crawford and Fabian (\cite{cra92}) is clearly visible
pointing outwards between $6\arcsec$ and $13\arcsec$ west of the center.
If the image is smoothed faint emission becomes evident all the way from 
the center to the filament. Other faint filaments  become visible as well,
e.g. one associated with the blue object seen in the contours  in the upper
left part of the figure
            }
  \label{emission}
\end{figure}

Optical imaging and spectroscopy were performed September 1996 using 
the DFOSC instrument
on the Danish 1.54m telescope at La~Silla. The field around the central cD 
galaxy is shown in Fig.~\ref{field}. The image was obtained by adding 
exposures in B (45 min), V (30 min), and Gunn\,I (30 min). 
The distribution of B-I colour for a $70\arcsec \times 70\arcsec$ area 
covering the central parts of the dominant cD galaxy is given in 
Fig.~\ref{B_I}. 
In order to image the distribution of the nebular emission we obtained narrow
band exposures through a filter ($\lambda 6908$, FWHM\ =\,98\,{\AA}, 1 hour)
covering the redshifted H$\alpha$+[N\,{\sc ii}] lines and an off-band filter
($\lambda 6801$, FWHM\,=\,98\,{\AA}, 1 hour). After scaling and subtraction 
a H$\alpha$+[N\,{\sc ii}] image is obtained. The central part of this image is
shown in Fig.~\ref{emission}.

Details about the spectroscopy are found in Table~\ref{spectro}. The slit
was positioned on the cD nucleus with two different position angles. 
P.A.=$270\degr$ covers the western filament, and P.A.=$21\degr$ passes
the object in the upper left corner of Fig.~\ref{emission} and covers the
jet-like emission to the northeast and southwest.

\section{Results}
\label{results}
The general brightness distribution in the maps is described most easily
for the C200 maps. The $135\mu {\rm m} $ and $200\mu {\rm m} $ maps
are rather similar. An enhancement is seen at the center in all four maps
concordant with the position of the cD. A maximum is present in the upper left 
corners. After rotating the revolution 200 maps $15 \degr$ into coincidence 
with the rev.~173 maps we find these maxima to overlap suggesting the presence
of one or more real sources. Similarly there are maxima in the upper right 
corners. Their positions and relative brightness in the maps can be
understood if a source is present in the upper right corner of the rev.~200
maps, but just outside the rev.~173 field. A third characteristic feature
is the brightness minimum to the lower left (i.e. south) of the center of 
the C200 maps. Again, when we compare the maps after rotation the reality
of this minimum is confirmed. We conclude that the brightness distribution seen 
in the C200 maps is real.

The C100 maps have the advantage of better resolution which improves the
possibility of identifying optical counterparts. However,
the reality of the peaks in the $100\mu {\rm m} $ maps is not convincing
when the maps are compared after rotation. Generally the peaks occur at
different locations. Even the central source is doubtful: The rev.~200
map shows a weak enhancement slightly displaced to the right of the center,
but the rev.~173 map shows a minimum at the same location. 

A comparison between the $60\mu {\rm m} $ maps is more successful. Both 
show a central enhancement (C100-1) although slightly displaced to the 
right (north) in the rev.~173 map. The maximum brightness (object C100-2) 
occurs in both maps
near the upper left corners and overlap after rotation. In Fig.~\ref{field}
the approximate positions of overlap is marked by numbers for the off-center
sources.  The rev.~200 map has a peak (C100-3) in the upper 
right corner which may be related to the source  present in the C200 
maps. Furthermore, the peak (C100-4) in the right part of the 
rev.~200 $60\mu {\rm m} $ map overlaps with an enhancement in the 
rev.~173 map. There are disagreements as well, however. The peak
obvious in the rev.~173 map below C100-4 (confirmed by the LSQ reductions) is
not visible in the rev.~200 map. We conclude that the $60\mu {\rm m} $
sources C100-1, C100-2, C100-3, and C100-4 are likely to be real, but that 
the present reduction software still produces artifacts calling for 
caution in the interpretation.

In paper{\sc \,i} we found that aperture photometry of the faint sources 
suffers significantly from the uncertainty in
the evaluation of the background level. We therefore prefer to position,
scale and subtract the PSF from the maps. The success in removing the
source is then evaluated by eye. By varying the scaling we estimate the
maximum and minimum acceptable flux. The median and its deviation from the 
limits are given in Table~\ref{fluxes} for our identified infrared sources.
We assume that the two sources in the upper corners of the C200 maps are 
identical to C100-2 and C100-3. The reality of C100-1 at $100\mu {\rm m} $ 
may be questionable. C100-3 is outside the field in the rev.~173 map. 
\begin{table*}
\caption[]{
Source fluxes determined from the PIA images (Jy) by positioning, scaling, and
subtracting the PSF. The quoted uncertainties are {\em not} statistical, but
are subjectively evaluated limits
}
\begin{flushleft}
\begin{tabular}{l l l l l l }
\hline
object & 
$60\mu{\rm m}$ & $100\mu{\rm m}$ & $135\mu{\rm m}$ & $200\mu{\rm m}$ & rev. \\
\hline
C100-1 & $0.05\pm0.03$ & $~~~~~~-~~~~$ & $0.05\pm0.04$ & $0.07\pm0.04$ & 173 \\
       & $0.05\pm0.02$ & $0.04\pm0.02$ & $0.07\pm0.05$ & $0.11\pm0.05$ & 200 \\
\hline
C100-2 & $0.22\pm0.04$ & $~~~~~~-~~~~$ & $0.22\pm0.06$ & $0.07\pm0.04$ & 173 \\
       & $0.27\pm0.06$ & $~~~~~~-~~~~$ & $0.30\pm0.10$ & $0.12\pm0.05$ & 200 \\
\hline
C100-3 & $~~~~~~-~~~~$ & $~~~~~~-~~~~$ & $~~~~~~-~~~~$ & $~~~~~~-~~~~$ & 173 \\
       & $0.12\pm0.05$ & $~~~~~~-~~~~$ & $0.25\pm0.05$ & $0.12\pm0.05$ & 200 \\
\hline
C100-4 & $0.11\pm0.04$ & $~~~~~~-~~~~$ & $~~~~~~-~~~~$ & $~~~~~~-~~~~$ & 173 \\
       & $0.10\pm0.03$ & $~~~~~~-~~~~$ & $~~~~~~-~~~~$ & $~~~~~~-~~~~$ & 200 \\
\hline
\end{tabular}
\end{flushleft}
\label{fluxes}
\end{table*}

\section{Discussion}
\label{discussion}
\subsection{The cD galaxy}
\label{cD}
The central infrared source, C100-1, is detected in all maps except at 
$100\mu {\rm m} $. The measured fluxes in the two independent observations
also agree within the limits. We therefore regard the source as real. 
A comparison with the list of Jura et al. (\cite{jur87}) shows that the
luminosity of S\'{e}rsic\,159-03 at $60\mu{\rm m}$ is larger than other 
early type galaxies detected by IRAS by an order of magnitude or more,
except the extraordinarily bright  galaxy NGC\,1275 which is the center of the
Perseus cluster cooling flow, and which is undergoing an encounter 
with an other galaxy (e.g. N{\o}rgaard-Nielsen et al., \cite{noe93}).

In a previous paper (Hansen~et~al.~\cite{han95}) we presented a model for the
infrared emission from Hydra~A measured by IRAS. We assumed that most of the
mass cooling out of the cluster gas ends up in low mass stars forming in the
flow. We further assumed that dust grains were able to grow in the cool 
pre-stellar clouds converting a fraction $y$ of the mass into grains. If
the mechanism is effective we expect $y \approx 1$\%. After a star has formed 
the remaining material is dispersed in the hot cluster gas. If a fraction $f$ 
is recycled to the hot phase a dust mass of $y \times f \times {\rm \dot{M}}$ 
is continuously injected into the cluster gas. At forehand we expect $f$ to
be approximately $1-50$\%. The grains are destroyed by sputtering on a time 
scale $\tau_{\rm d}$, and a steady state is obtained. At any time a dust mass 
of ${\rm M_{d}}\,=~y \times f \times {\rm \dot{M}} \times \tau_{\rm d}$
is present. The grains are heated by hot electrons (in the inner galaxy the
photon field may also be important), and the infrared emission can be 
evaluated. The present data do not  allow testing of more elaborate
models having radial distributions of e.g. the dust temperature. We therefore
only make a simple estimate using mean values.

For Hydra~A we found that $y\,=~1\%$ and $f\,=~11\%$ reproduced the observed
IRAS flux. In Table~\ref{model} giving calculated fluxes we repeat the 
calculations for S\'{e}rsic\,159-03, but with $f$ reduced to 2\%. 
Considering the crude model and the uncertainty of the measurements we find
the agreement with the observed values in Table~\ref{fluxes} satisfactory.
This result has some significance although $f$ has been used as a free
parameter to obtain concordance. If a value of $f$ much larger than unity
had been necessary to fit the observations the model would have had to be 
rejected. Also, a value significantly lower than 1\% would have made the 
model unconvincing.
\begin{table*}
\caption[]{
Mass deposition rate, temperature and cooling radius for the cluster gas 
(Allen and Fabian \cite{all97}). Predicted infrared fluxes from dust grains
are given for a simple model. The model assumes the presence of dust in
the gas as a by-product of star formation in the flow. The calculated dust
temperature and total dust mass are also given
}
\begin{flushleft}
\begin{tabular}{l l l l l l l l l}
\hline
${\rm \dot{M}}$ & ${\rm T_{x-ray}}$ & ${\rm R_{cool}}$ &
$60\mu{\rm m}$ & $100\mu{\rm m}$ & $135\mu{\rm m}$ & $200\mu{\rm m}$ &
${\rm T_{dust}}$ & ${\rm M_{dust}}$  \\
${\rm M_{\sun} ~yr^{-1}}$ & keV & arcmin & Jy & Jy & Jy & Jy & 
K & ${\rm M_{\sun}}$ \\
\hline
231  & 2.9 & 1.89  & 0.05  & 0.07  & 0.06  & 0.03 & 40 & $1.4\,10^{6}$ \\
\hline
\end{tabular}
\end{flushleft}
\label{model}
\end{table*}

A possible disagreement with the model is, however, the small extent of the
source. One would expect the infrared emission to show some distribution
within the cooling radius which is $1 \farcm 89$. Although the resolution at
$60\mu{\rm m}$ is $50 \arcsec$ C100-1 is indistinguishable from a point source
in all our measurements. The reason could be that (1) the star formation
is concentrated to the center (as seems to be the case for Hydra~A, see Hansen
et al.~\cite{han95}), (2) the model does not apply, or (3) instrumental
effects prevents detection of a faint, extended distribution of FIR
emission. 

Alternative possibilities are that C100-1 is related to the active nucleus
as inferred by the presence of a radio source (Large et al. \cite{lar81},
Wright et al. \cite{wri94}), or that dust has been introduced into the 
system by a recent merger event. A hint may be that all three measurable
images of the revolution 173 maps show a tendency to be displaced from the
center by $\approx 10 \arcsec$ to the north where  nebular line emission 
is seen (Fig.~\ref{emission}). The cD galaxy shows no signs of dust lanes, but
exhibits a constant distribution in colour (Fig.~\ref{B_I}). There are,
however, two objects in the upper left part of Fig.~\ref{B_I} which are
bluer than the cD. The brightest and bluest of these looks disturbed
possibly due to tidal interaction. The spectra taken with P.A.\,=~$21\degr$
cover the object and contain emission lines. The emission is weak in
Fig.~\ref{emission} because the lines are shifted away from the peak 
transmission of the filter. Relative to the cD we find the velocity of the
object to be $+1800\pm 200\,{\rm km~s^{-1}}$. The galaxy may have plumped
through the cD and contains young stars.

The origin of the optical filaments in Fig.~\ref{emission} is a puzzle.
It may be captured material from mergers, related to radio plasma, or
connected to the cooling flow. The relative velocities do not support any
particular model. The velocities have been measured from our spectra, and 
they are quite low as seen from Table~\ref{velo}. 
\begin{table*}
\caption[]{
Velocities of optical filaments relative to the nuclear \\
emission of the cD galaxy (z\,=~0.0568)
          }
\begin{flushleft}
\begin{tabular}{l c c}
\hline
filament      &  rel. vel.  \\
\hline
P.A. $21\degr$ north  & $-30\pm30$  & ${\rm km~s^{-1}}$  \\
P.A. $21\degr$ south  & $-110\pm60$ &   --       \\
P.A. $270\degr$ west  & $-120\pm20$ &   --       \\
\hline
\end{tabular}
\end{flushleft}
\label{velo}
\end{table*}
Donahue and Voit (\cite{don93}) obtained spectra of the nuclear emission 
from the S\'{e}rsic\,159-03 cD galaxy. They argued that the lack of 
[Ca\,{\sc ii}] $\lambda 7291$ emission indicates that Ca is depleted onto 
dust grains. We have added all our spectra of the center together and all of 
the filaments. No [Ca\,{\sc ii}]  emission was visible in any of the two
resulting spectra. We then shifted the [N\,{\sc ii}]$\lambda 6583$ to
the expected position of [Ca\,{\sc ii}] and added the shifted line after
scaling with various constants. In this way we find that no [Ca\,{\sc ii}]
emission stronger than 0.20 times [N\,{\sc ii}]$\lambda 6583$ is present.
Figure~1 of Donahue and Voit (\cite{don93}) predicts (from ionization
calculations) that this ratio should 
never be smaller than 0.24. Although marginal compared to the case of
Hydra~A the discrepancy can be explained by the 
condensation of Ca onto grains in accord with Donahue and Voit's result.

The presence of dust in the nebular gas does not necessarily exclude 
that it originates from the cooling cluster gas. Dust may grow in dense, cool
clouds in connection with star formation. For the nebular gas in Hydra~A
Donahue and Voit (\cite{don93}) found a much tighter limit on the 
[Ca\,{\sc ii}] line strongly suggesting the presence of dust. In Hydra~A
the nebular gas is concentrated to a central disk-like structure of
several kpc where vigorous star formation has taken place, and
Hansen~et~al.~(\cite{han95}) argue that it is a result of the cooling flow
(see also McNamara, \cite{mcn95}). In S\'{e}rsic\,159-03 the extended nature 
of the filaments and the presence of the blue, star forming object is more 
in favour of a merger scenario, however.

\subsection{Off-center infrared sources}
\label{non_central}
There are no striking optical identifications to the off-center sources.
The position of C100-2 is relatively well determined by the overlap of the
two observations. The nearest object visible in Fig.~\ref{field} is 
$\approx 0 \farcm 5 $ to the south-west, is unresolved 
and of blue colour. It is not a known QSO (no QSO is closer than 
$30 \arcmin$ in the NASA/IPAC Extragalactic Database\footnote {The 
NASA/IPAC Extragalactic Database (NED) is operated by the Jet Propulsion 
Laboratory, California Institute of Technology, under contract with the 
National Aeronautics and Space Administration }),
and it is just outside the overlap of the two observations. There are several 
faint optical objects in the area of C100-3, but no show up in our data with 
characteristics
favouring a candidateship. The difficulties in pointing out candidates
are even more pronounced for C100-4 which agrees poorly with the nearest faint
objects in Fig.~\ref{field}. However, C100-4 is also the most uncertain of
the sources as it is only visible at $60\mu {\rm m} $.

\section{Conclusion}
\label{conclusion}
The availability of two observations covering essentially the same field
at several wavelengths allows us to identify 4 faint ($\approx 0.1$~Jy)
far-infrared sources with some confidence. A central source, C100-1, is 
attributed to the cD galaxy which contains optical filaments, but our
optical images do not reveal significant evidence of dust lanes. 
The fluxes measured for C100-1 are of the same
order of magnitude as expected from dust related to star formation in
the cooling flow. For the non-central sources we cannot point out any 
particular optical candidates
in contrast to the results from the Abell\,2670 field (paper{\sc \,i}) where
galaxies with enhanced star formation were found coincident with the
infrared sources.

\acknowledgements
{This work has been supported by The Danish Board for Astronomical Research.
}


\begin{thebibliography}{}
%
  \bibitem[1989]{abe89}
    Abell G.O., Corwin H.G., Olowin R.P., 1989, ApJS 70, 1
  \bibitem[1997]{all97}
    Allen S.W., Fabian A.C., 1997, MNRAS 286, 583
  \bibitem[1992]{cra92}
    Crawford C.S., Fabian A.C., 1992, MNRAS 259, 265
  \bibitem[1993]{cra93}
    Crawford C.S., Fabian A.C., 1993, MNRAS 265, 431
  \bibitem[1993]{don93}
    Donahue M., Voit G.M., 1993, ApJ 414, L17
  \bibitem[1995]{han95}
    Hansen L., J{\o}rgensen H.E., N{\o}rgaard-Nielsen H.U., 1995, A\&A 297, 13
  \bibitem[1999]{han99}
    Hansen L., J{\o}rgensen H.E., N{\o}rgaard-Nielsen H.U., Pedersen K.,
    Goudfrooij P., Linden-V{\o}rnle M.J.D., 1999,
    A\&A 349, 406 (paper{\sc \,i})
  \bibitem[1989]{hec89}
    Heckman T.M., Baum S.A., van Breugel W.J.M., McCarthy P., 1989, ApJ 338, 48
  \bibitem[1987]{jur87}
    Jura M., Kim D.W., Knapp G.R., Guhathakurta P., 1987, ApJ 312, L11
  \bibitem[1981]{lar81}
    Large M.I., Mills B.Y., Little A.G., Crawford D.F., Sutton J.M., 
    1981, MNRAS 194, 693
  \bibitem[1996]{lem96}
    Lemke D. et al., 1996, A\&A 315, L64
  \bibitem[1995]{mcn95}
    McNamara B.R., 1995, ApJ 443, 77
  \bibitem[1993]{noe93}
    N{\o}rgaard-Nielsen H.U., Goudfrooij P., J{\o}rgensen H.E., Hansen L.,
    1993, A\&A 279, 61
  \bibitem[1994]{wri94}
    Wright A.E., Griffith M.R., Burke B.F., Ekers R.D., 1994, ApJS 91, 111
%
\end{thebibliography}
\end{document}